\documentclass[12pt]{article}
\usepackage{amsmath,amssymb}
\usepackage{graphicx,color}
\numberwithin{equation}{section}
\usepackage{cite}
\usepackage{bm}
\usepackage{dcolumn}
\newcommand{\be}{\begin{equation}}
\newcommand{\bea}{\begin{eqnarray}}
\newcommand{\eea}{\end{eqnarray}}
\newcommand{\ba}{\begin{array}}
\newcommand{\ea}{\end{array}}
\newcommand{\ee}{\end{equation}}

\expandafter\ifx\csname mathbbm\endcsname\relax

\newcommand{\ds}{\!\!\!\!\! \hspace{1mm} /}
\else

\fi \textheight 22cm \textwidth 15cm \topmargin 1mm
\begin{document}

\begin{titlepage}
\hfill
\vbox{
    \halign{#\hfil         \cr
           IPM/P-2011/034 \cr
                      } 
      }  
\vspace*{20mm}
\begin{center}
{\Large {\bf Boundary CFT from Holography}\\
}

\vspace*{15mm}
\vspace*{1mm}
{Mohsen Alishahiha and  Reza Fareghbal  }

 \vspace*{1cm}

{\it  School of physics, Institute for Research in Fundamental Sciences (IPM)\\
P.O. Box 19395-5531, Tehran, Iran \\ }

\vspace*{.4cm}

{E-mails: {\tt alishah@ipm.ir, fareghbal@theory.ipm.ac.ir}}%

\vspace*{2cm}
\end{center}

\begin{abstract}

We explore  some aspects of holographic dual of  Boundary Conformal
Field Theory (BCFT). In particular we study asymptotic symmetry of
geometries which provide holographic dual of BCFTs. We also compute
two-point functions of certain bosonic and fermionic operators in
the dual BCFT by making use of AdS/BCFT correspondence.

\end{abstract}

\end{titlepage}

\section{Introduction}

Extension of AdS/CFT correspondence
\cite{{Maldacena:1997re},{Gubser:1998bc},{Witten:1998qj}} to the
case of boundary conformal field theories (BCFTs) has recently been
addressed in \cite{{Takayanagi:2011zk},{Fujita:2011fp}}.
The idea was to start with the standard AdS/CFT correspondence and to seek for a
modification of the bulk gravitational theory such that the dual
theory becomes a CFT defined on a space with a boundary.

Actually, the main idea of the AdS/BCFT construction of \cite{Takayanagi:2011zk} is as follows:
 One may start with an asymptotically locally AdS geometry where we typically
impose Dirichlet boundary on the metric as we approach the boundary.
It is, however, possible to modify the geometry by imposing two
different boundary conditions on the metric as one approaches the
boundary. This procedure automatically implies  that the boundary is
divided
 into two parts. While in the first part, where the BCFT is
supposed to live, we still impose the Dirichlet boundary condition;
on the other part, the metric would satisfy the Neumann boundary
condition.  The interface between two parts is, indeed, the boundary
of the space where the BCFT is defined.

\begin{figure}
\begin{center}
\includegraphics[height=6cm, width=9cm]{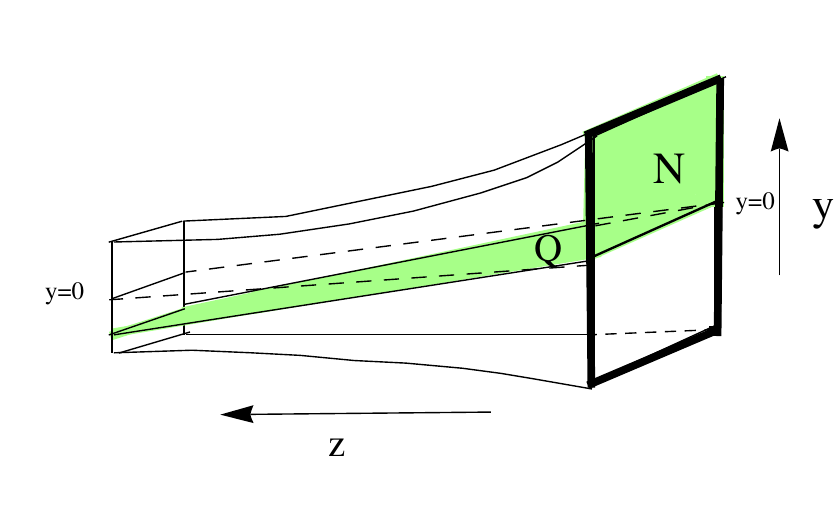}
\caption{AdS geometry with a new boundary $Q$, which divides the
space into two parts.
 While we impose the Dirichlet
boundary condition on $N$, on $Q$ the metric satisfies the  Neumann boundary condition.
The geometry  provides holographic dual of a BCFT in half space. The boundary $Q$
is given by equations \eqref{QS}.} \label{fig1}
\end{center}
\end{figure}

More precisely, consider a gravitational model which admits an AdS
vacuum solution. The simplest action contains Einstein-Hilbert
action with a negative cosmological constant \be S=\frac{1}{16\pi
G_N}\int_M d^{d+1}x\sqrt{-g}(R-2\Lambda)+\frac{1}{8\pi
G_N}\int_{\partial M} d^dx\sqrt{-h} K+S_{\rm matter}, \ee where $g$
and $h$ are the bulk and boundary metrics, respectively. The second
term is the Gibbons-Hawking boundary term \cite{Gibbons:1976ue}
which is given by the trace of extrinsic curvature $K=h^{ab}K_{ab}$.
$S_{\rm matter}$ is a possible matter field one may add on the
boundary. Following the above construction, the boundary is divided
into two parts $\partial M=N\cup Q$ such that $\partial Q=\partial
N$.  The metric satisfies the Neumann boundary condition on $Q$.
With this boundary condition the bulk geometry would also be
modified in such a way that the gravitational theory lives  only in
a portion of the whole AdS space (see, for example, figure
\ref{fig1}) . The modified geometry would provide a holographic dual
for BCFT.

The variation of the action with respect to the metric leads to the
following boundary terms \bea \delta S&=&\frac{1}{16\pi G_N} \int_N
d^dx\sqrt{-h}(K_{ab}-Kh_{ab})\delta h^{ab}+ \frac{1}{16\pi G_N}
\int_Q d^dx\sqrt{-h}(K_{ab}-Kh_{ab})\delta h^{ab}\cr &&\cr &&
\;\;\;\;\;\;\;\;\;\;\;\;\;\;\;\;\;\;\;+\frac{1}{2}\int_Q
d^dx\;\sqrt{-h}\;T_{ab}\;\delta h^{ab}. \eea Here the last term
came from the variation of the  matter action  defined
 on the boundary $Q$.
 While the
metric satisfies the Dirichlet boundary condition on the subboundary
$N$, we impose the Neumann boundary condition on subspace $Q$, which
leads to the following constraint on the metric
\cite{Takayanagi:2011zk}:
\be\label{neumann-condition} K_{ab}-K
h_{ab}=8\pi G_N T_{ab}.
 \ee
 When we add a constant  matter field, the above equation reads
\be\label{BC} K_{ab}=(K-T)h_{ab}, \ee where the constant $T$ may be
interpreted as the tension of the boundary surface $Q$
\cite{Takayanagi:2011zk}.

In practice, one may start with an asymptotically locally AdS$_{d+1}$
solution parameterized by $(z,y, x_1\cdots,x_{d-1})$ and then using the
boundary condition \eqref{BC}, the boundary surface $Q$ may be
described by a hypersurface given by a curve $f(z,y,\cdots,x_{d-1})=0$. Although we could
proceed with a complicated boundary, in what follows we will
consider the simplest case where the CFT  lives in a half space.
More precisely we will consider the case where the BCFT lives on a
$d$ dimensional space parameterized by $(y,x_1,\cdots,x_{d-1})$ for
$y\geq 0$. In other words $y$  denotes the perpendicular distance
from the boundary defined at $y=0$.

Let us consider an  AdS$_{d+1}$ geometry in the Poincar\'e
coordinates\footnote{The radius of the AdS is set to one.}
\be\label{met} ds^2=\frac{dz^2+dy^2-dx_1^2+\cdots+dx_{d-1}}{z^2}.
\ee Then the boundary $Q$   is given by the following  curve which
is, indeed, a solution of the boundary condition \eqref{BC}
\be\label{QS} y(z)=\frac{T z}{\sqrt{(d-1)^2-T^2}}. \ee With this
boundary the AdS geometry is divided into two parts where the
gravitational theory lives in the upper half of the geometry as
depicted in figure \ref{fig1}. This geometry provides a holographic
description of a BCFT on the half space defined by
$(y,x_1,\cdots,x_{d-1})$ with $y>0$. This is the model we will be
mostly studying in this letter.

It is the aim of this letter to further explore  some aspects of
AdS/BCFT correspondence. In particular we study asymptotic symmetry
of geometries which provide holographic dual of BCFT. We also
compute the two-point function of certain bosonic and fermionic
operators in the BCFT, using AdS/BCFT correspondence. The resultant
two-point functions agree with those in the literature of BCFT.
Therefore this may be thought of as a nontrivial check for recently
conjectured AdS/BCFT correspondence.

The letter is organized as follows: In the next section we shall
study the asymptotic symmetry of geometries which provide
holographic dual of BCFTs. In section three utilizing  the
holographic description, we compute the two-point function of a scalar
operator in the dual BCFT. In section four we will redo the same
computations for a fermionic operator.
The last section is devoted to  discussion.

\section{Asymptotic symmetry}

In this section we would like to study  asymptotic symmetry of
geometries which provide holographic dual of BCFTs.  Following the
general sprite of the AdS/CFT correspondence, one expects that the
corresponding asymptotic symmetry will be the symmetry of  BCFTs.

\begin{figure}
\begin{center}
\includegraphics[height=6cm, width=9cm]{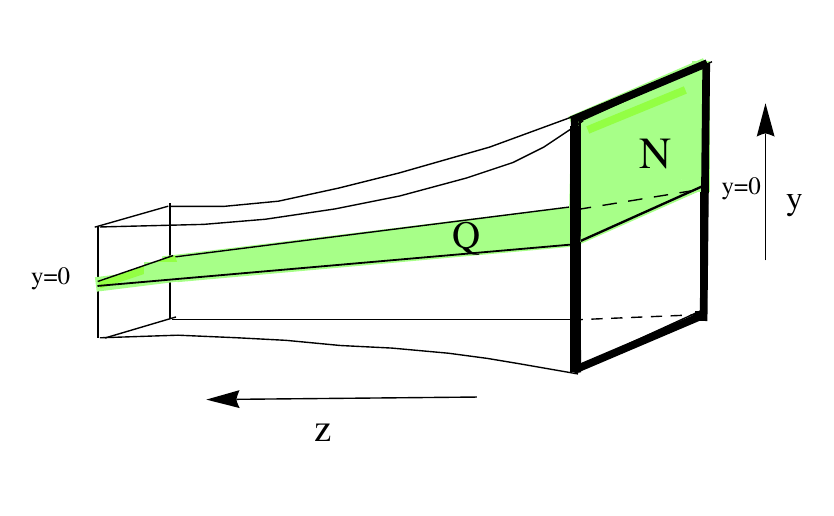}
\caption{ AdS geometry with tensionless boundary $Q$ at $y=0$.} \label{fig2}
\end{center}
\end{figure}

Although we can explore the asymptotic symmetry of a generic BCFT,
we will address the question for the case where the dual BCFT lives
in  a half space. The  corresponding geometry is, indeed, given in
the previous section. Moreover for simplicity we will also assume
that the boundary $Q$ has zero tension, {\it i.e.} $T=0$. In this
case, the boundary $Q$ is defined by the hypersurface $y=0$ as shown
in figure \ref{fig2}.

We note that, although the asymptotic symmetry can be worked out for
any dimensions, the interesting case would be a two dimensional
 BCFT where one expects that the symmetry enhances to a
Virasoro algebra. Therefore, in what follows we will consider the
two dimensional BCFTs.

To proceed, it is useful to define coordinates $x^\pm=x_1\pm y$ in
which the metric \eqref{met} becomes \be
ds^2=\frac{dz^2-dx^+dx^-}{z^2}, \ee and the boundary is defined by
$x^+=x^-$. This would provide a holographic dual for a BCFT defined
in the upper half plane.

To study the asymptotic symmetry, we perturb the above metric such
that the fall off of the components of the resultant geometry
satisfy the following boundary conditions\cite{Brown:1986nw}
\be \delta g_{zz}=\delta
g_{z+}=\delta g_{z-}={\mathcal{O}}(z),\qquad \delta
g_{+-}={\mathcal{O}}(1) ,
\ee
as one approaches the boundary at $z=0$.
The asymptotic Killing vectors which leave the above conditions
unchanged are given by\cite{Brown:1986nw}
\begin{eqnarray}\label{generators of ASG}
  \zeta^z &=& {z\over2}\left[{dT^+(x^+)\over dx^+}+{dT^-(x^-)\over dx^-}\right]+\cdots\; ,\cr
  \zeta^+ &=& T^+(x^+)+{z^2\over2}{d^2T^-(x^-)\over dx^{-\,2}}+\cdots\;, \cr
  \zeta^- &=& T^-(x^-)+{z^2\over2}{d^2T^(x^+)\over dx^{+\,2}}+\cdots\;,
\end{eqnarray}
where $T^+,T^-$ are arbitrary functions  of $x^+$ and $x^-$, respectively. Expanding these  functions in the form of
\be
\label{expansion} T^\pm(x^\pm)=\sum_n L_n^\pm e^{inx^\pm},
\ee
results in two  Virasoro algebras generated by  $L^+_n$ and $L^-_n$.

Actually, since the above considerations are based on the tensorial
relations, imposing the  Neumann condition \eqref{neumann-condition}
does not impose any constraint on the generators given in
\eqref{generators of ASG}. We note, however, that in general the
above asymptotic Killing vectors may change the location of the
boundary. More precisely, under the action of the above asymptotic
Killing vectors, in leading order, $x^\pm$ maps into ${\tilde
x}^\pm$ as follows \bea\label{xpm} {\tilde x}^+ =x^+
+T^+(x^+)+{z^2\over2}{d^2T^-(x^-)\over dx^{-\,2}}, \;\;\;\;\;\;
 {\tilde x}^- = x^-+T^-(x^-)+{z^2\over2}{d^2T^(x^+)\over dx^{+\,2}},
\eea It is clear from \eqref{xpm} that the boundary defined by
$x^+=x^-$, will not remain fixed  by the generators of the asymptotic
symmetry group. In order to keep it fixed, one needs to impose the
following condition on the generators of the asymptotic Killing
vectors
\be \label{condition}
T^+(x^+)\big|_{x^+=x^-}=T^-(x^-)\big|_{x^+=x^-}.
\ee
By making use
of \eqref{expansion} one finds that  $L^+_n=L^-_n$  and thus
the functions $T^\pm(x^\pm)$ are no longer independent. As a result, we find that the asymptotic
symmetry of geometries which provides a holographic dual of  BCFTs  is effectively a copy
of the Virasoro algebra. More precisely, the left and right moving modes
are related, as expected.

We should mention that  asymptotic symmetries of a theory of gravity in a background consisting of two patches of AdS3 spacetime glued together along an AdS$_2$ brane
has been studied in
\cite{Bachas:2002nz} .  The corresponding symmetry is generated by a single Virasoro algebra.
We note that this is essentially the asymptotic symmetry of a geometry which could
provide holographic dual for BCFT.

It is worth mentioning that the simple relation   we have
found between $L^{+}_n$ and $L^-_n$, $(L^+_n=L^-_n)$, is a consequence of the simple boundary  we have chosen, {\it i.e.} $y=0$.
Had we considered a more complicated boundary, $L^+_n$ would be related to $L^-_n$
in a more involved relation.

It is easy to generalize the above consideration to higher
dimensions. Doing so, one finds that the asymptotic symmetry group
is $SO(1,d)$ which is the symmetry group of Euclidean d-dimensional BCFT.

\section{Correlation functions of scalar field  }

Having discussed the holographic  dual of BCFTs, it would be
interesting to study correlation  functions of different operators in the BCFT by
making use of its holographic dual.  Indeed, it is the aim of this
section to evaluate the two-point function of a bosonic operator in an Euclidean  BCFT
in half space using
its  holographic dual. The corresponding  bulk geometry, given in the previous section,
 is\footnote{We use a notation in which
 $\vec {X}=(y,\vec{x})\equiv(y,x_i)=(y,x_1,\cdots,x_{d-1})$.}
\be\label{backg} ds^2={1\over z^2}\left(dz^2+dy^2+d\vec
x\,^2\right)\qquad y,z\geq 0 \ee which provides a holographic dual for
the  $d$ dimensional BCFT in the half space parameterized by
$(y,x_1,\cdots,x_{d-1})$  with $y\geq 0$. We recall that by
construction, the above metric has two boundaries at $z=0$ and $y=0$;
while we impose the Dirichlet boundary condition on the metric at
$z=0$, the metric satisfies the  Neumann condition on $y=0$.

Let us consider a massive free scalar field in the background \eqref{backg}
\be\label{ac}
S={1\over2}\int\,d^{d+1}x\,\sqrt{g}\left(\partial_\mu\Phi\partial^\mu\Phi+m^2\Phi^2\right).
\ee
We would like to calculate the two-point function of the corresponding
dual operator using the general rules of AdS/CFT correspondence.
To do so, we
will mostly follow the notation of \cite{Skenderis:2002wp}.

The linear variation of the action \eqref{ac} leads to the following boundary
term\footnote{ We note that in the present case where we have two boundaries, $Q$ and $N$,
it might be necessary to add another boundary term on $Q$  in order to maintain conformal
symmetry on the boundary $N$ (to have a BCFT on $N$). Indeed a boundary term  might be
crucial to prevent energy flow on the boundary $Q$. We would like to thank M. M. Sheikh Jabbari for a  comment on this point.}
\be
\delta S_{\rm{ bdy}}=\int_{\partial M}d^dx \sqrt{h}\;
n^\mu\delta\Phi\partial_\mu\Phi,
\ee
 where $n^\mu$ is a unit vector
normal to the boundary. Since in our case  the boundary is made of
two parts $\partial M=N\cup Q$ defined by $z=0$ and $y=0$ surfaces,
respectively, one arrives at
\be
\delta S_{\rm{ bdy}}=\int_{N}d^dx
\sqrt{g}\;g^{zz} \delta\Phi\partial_z\Phi+ \int_{Q}d^dx
\sqrt{g}\;g^{yy} \delta\Phi\partial_y\Phi.
\ee
The boundary
condition at $z=0$ follows from the standard dictionary of AdS/CFT
correspondence. Usually we impose the Dirichlet boundary condition
at $z=0$.  Actually, in general, the asymptotic behavior of the
scalar field as one approaches $z=0$ may be recast to the following
form \be\label{expansion of scalar} \Phi(z,\vec
X)=z^{d-\Delta}[\phi_{(0)}+z^2\phi_{(2)}+\cdots+z^{2\Delta-d}\left(\phi_{(2\Delta-d)}+
2\psi_{(2\Delta-d)}\ln(z)\right)+\cdots] ,
\ee
 where  $\Delta$ is
determined through equation $m^2-\Delta(\Delta-d)=0.$

According to the dictionary of the AdS/CFT correspondence, $\phi_0$
is interpreted as the source of a dual scalar operator,
$\mathcal{O}$, with scaling dimension $\Delta$ in the dual conformal
field theory. Moreover, the expectation value of the dual operator,
$\langle \mathcal{O} \rangle$, is determined by $\phi_{(2\Delta-d)}$
up to local counterterms. The corresponding two-point function
is also given by
\be\label{2point}
\langle\mathcal{O}\mathcal{O}\rangle=
-{\delta\phi_{(2\Delta-d)}\over
\delta\phi_{(0)}}\big
|_{\phi_{(0)}=0}.
\ee

On the other hand, one should also impose a proper boundary condition
on the other boundary, $Q$, at  $y=0$. In this case,  one may impose
either Dirichlet, $\Phi|_{y=0}=0$, or  Neumann,
$\partial_y\Phi|_{y=0}=0$, boundary conditions.
Therefore,  in order to  find the two-point function of scalar operator $\mathcal{O}$,
 one needs to solve the equation of motion with  the above  boundary conditions.

To proceed, we start with  an ansatz \be\Phi( \vec
x,y,z)=z^{d\over2}f(z)h(y)\exp(-i\,\vec \omega.\vec x), \ee by which
the equation of motion reads \be {1\over
z^2f(z)}\left(z^2{d^2f(z)\over dz^2}+z{d f(z)\over dz}-(\nu^2+
\vec\omega\,^2 z^2)f(z)\right)+{1\over h(y)}{d^2h(y)\over dy^2}=0,
\ee where $\nu=\Delta-{d\over2}$. Therefore, one finds
 \be\label{eq1}
  z^2{d^2f(z)\over dz^2}+z{d f(z)\over dz}=\left(\nu^2+
 k^2 z^2\right)f(z)
\ee
and
 \be\label{eq2}
  {d^2h(y)\over dy^2} = -q^2 h(y)
\ee
 where $q$ is a constant and $k^2=\vec\omega\,^2+q^2$.

It is easy to solve these equations. In particular, for the Dirichlet
or Neumann boundary conditions the solutions of the equation
\eqref{eq2} are
 \be h(y)=c_0(e^{-iqy}\pm e^{iqy}),
\ee
 where $c_0$
is a constant and the plus sign is for the Neumann boundary
condition while the minus sign is for the Dirichlet one.

Therefore, the most general solution for $\phi$ may be written as
follows
\be\label{phi}
\Phi( \vec
x,y,z)=\frac{z^{d/2}}{ (2\pi)^d 2^{\nu}\Gamma(\nu)}\int_{-\infty}^{+\infty}\,d^{d}k
\;k^{\nu}K_\nu(k z)e^{-i\vec \omega \cdot \vec x}(e^{-iqy}\pm
e^{iqy})\phi_{(0)}(\vec\omega,q), \ee where $K_\nu(kz)$ is modified
Bessel function and $\phi_{(0)}(\vec\omega,q)$ is the source of the
dual operator. Note that in our notation $d^dk$ stands for
$d^{d-1}\omega dq$.  Moreover,  in order to normalize the source we have
explicitly put a factor of $k^{\nu}$.

An immediate consequence of the above expression for a general
solution of the equation of motion is that in order for the solution to
satisfy the desired boundary conditions, the source
$\phi_{(0)}(\vec\omega,q)$ should be either an even or odd function
with respect  to $q$.  More precisely, for
 Dirichlet condition
$\Phi|_{y=0}=0$ one finds  \be\label{condition D}
\phi_{(0)}(\vec\omega,-q)=-\phi_{(0)}(\vec\omega,q)
\ee
while for Neumann boundary condition $\partial_y\Phi|_{y=0}=0$ we get
\be\label{condition N}
\phi_{(0)}(\vec\omega,-q)=\phi_{(0)}(\vec\omega,q).
\ee
Therefore  we have
\be \phi_{(0)}(\vec\omega,q)=\frac{1}{2 (2\pi)^d}\int d^{d-1}x'dy'
e^{i\vec\omega .\vec x\,'}(e^{iqy'}\pm e^{-iqy'})\phi_{(0)}(\vec
x\,',y'), \ee where $\phi_{(0)}(\vec x\,',y')$ is the Fourier
transform  of the source $\phi_{(0)}(\vec\omega,q)$. Plugging this
expression into the equation \eqref{phi} we arrive at \be \Phi(\vec
x,y,z)=\int\,d^{d-1} x' dy'\phi_{(0)}(\vec x\,',y')G( \vec x,y;\vec
x\,',y',z) \ee where $G(\vec x,y;\vec x\,',y',z)$ is the
bulk-to-boundary  propagator of the scalar field in the present of the
boundary  at $y$ which can be expressed in terms of the
bulk-to-boundary  propagator of the scalar field when there is no
boundary, $G_0(\vec x,y;\vec x\,',y',z)$,  as follows:
\be\begin{split} G(\vec x,y;\vec x\,',y',z)={1\over4}\Big(G_0(\vec
x,y;\vec x\,',y',z)&\pm G_0(\vec x,-y;\vec x\,',y',z)\pm G_0(\vec
x,y;\vec x\,',-y',z)\cr &+G_0(\vec x,-y;\vec
x\,',-y',z)\Big)\end{split} \ee where the plus and minus signs  are
for the Neumann and
 Dirichlet boundary conditions, respectively. In our notation,
the bulk-to-boundary propagator for the scalar field on the $AdS$
space without a boundary, $G_0(\vec x,y;\vec x\,',y',z)$,  is
 \be
 G_0(\vec
x,y;\vec x\,',y',z)=\frac{z^{d/2}}{(2\pi)^d 2^{\nu}\Gamma(\nu)
}\int\, d^dk\,k^{\nu} K_\nu(k z)\;e^{i\vec\omega.(\vec x-\vec
x\,')}e^{iq(y-y')}.
 \ee
From this
expression and utilizing  the asymptotic behavior of modified Bessel
function, one can read the two-point function  by making use of the
general rule given by the equation \eqref{2point}. The resultant two-point
function is
\be\label{2pointFun } \langle{\mathcal{O}}(\vec
X_1){\mathcal{O}}(\vec X_2)\rangle_{CFT}\sim{1\over|\vec X_1- \vec
X_2|^{2\Delta}}. \ee

Therefore the corresponding  two function of the boundary CFT is found as follows
\be\label{hologhraphic 2point function}
\begin{split}\langle{\mathcal{O}}(\vec X_1){\mathcal{O}}(\vec X_2)\rangle_{BCFT}
={1\over4}&\Big(\langle{\mathcal{O}}(\vec X_1){\mathcal{O}}(\vec X_2)\rangle_{CFT}
\pm \langle{\mathcal{O}}(\vec X_1){\mathcal{O}}(\vec X\,^*_2)\rangle_{CFT}\cr &\pm
\langle{\mathcal{O}}(\vec X\,^*_1){\mathcal{O}}(\vec X_2)\rangle_{CFT}+
\langle{\mathcal{O}}(\vec X\,^*_1){\mathcal{O}}(\vec X\,^*_2)\rangle_{CFT}\Big)\end{split}
\ee
that is
\be\label{BCFT2point}
\langle{\mathcal{O}}(\vec X_1){\mathcal{O}}(\vec X_2)\rangle_{BCFT}\sim{1\over4}\Big({1\over|\vec X_1-\vec X_2|^{2\Delta}}\pm{1\over|\vec X_1-\vec X^*_2|^{2\Delta}}\pm
{1\over|\vec X^*_1-\vec X_2|^{2\Delta}} +{1\over|\vec X^*_1-\vec X^*_2|^{2\Delta}}\Big)
\ee
where $\vec X\,^*_{1,2}$  are images of  $\vec X\,_{1,2}$ with respect to the boundary, {\it i.e.}
if $\vec X=(y,\vec x)$ then  $\vec X^*=(-y,\vec x)$.
Defining
\be\zeta={|\vec X_1-\vec X_2||\vec X\,^*_1-\vec X\,^*_2|\over|\vec X_1-\vec X\,^*_1||
\vec X_2-\vec X\,^*_2|},
\ee
it is notable that  the  expression \eqref{hologhraphic 2point function}
may be written in the following form
\be \langle{\mathcal{O}}(\vec X_1){\mathcal{O}}(\vec X_2)\rangle_{BCFT}
=\left[{|\vec X_1-\vec X\,^*_1||\vec X_2-\vec X\,^*_2|\over |\vec X_1-\vec X_2|
|\vec X\,^*_1-\vec X\,^*_2||\vec X_1-\vec X\,^*_2||\vec X\,^*_1-\vec X_2|}\right]^\Delta F(\zeta),
 \ee
where \be
F(\zeta)=(constant)\left[(\zeta+1)^\Delta\pm\zeta^\Delta\right] \ee
Again the plus and minus signs correspond to the Neumann and
Dirichlet boundary conditions, respectively.  We note that the
resultant two-point function has the expected form of the
correlation function of scalar operators in
BCFTs\cite{Cardy:1984bb}. Therefore, one may want to conclude that
reproducing the expected results for the correlation function from
holographic dual could be thought of as a nontrivial check for the
recently proposed AdS/BCFT
correspondence\cite{{Takayanagi:2011zk},{Fujita:2011fp}}.


\section{Correlation functions of fermions}

In this section, we study the two point function of a fermionic operator
in the BCFT considered in the previous sections, using its gravity dual.
To do so, we start with  a fermionic field in the bulk whose action  is \be
S=\int_M d^{d+1}x\sqrt{g}\;\bar{\psi}\;( D\ds-m)\psi+\int_{\partial
M} d^dx \sqrt{h}\; \bar{\psi}\psi \ee As it was shown in
\cite{Henneaux:1998ch} the present of the boundary term is crucial
to
 get a well-defined variational principle. Actually,  this boundary term is also necessary for
AdS/CFT to work \cite{{Henningson:1998cd},{Mueck:1998iz}}
\footnote{See \cite{Laia:2011zn} for a nice description  of possible
boundary terms and boundary conditions.}. In what follows, we will
mostly follow the notation of \cite{Mueck:1998iz}.

The corresponding equations of motion are
\bea
&&(D\ds-m)\psi=\left(z\gamma_\mu\partial_\mu -\frac{d}{2}\gamma_z-m\right)\psi=0,\cr
&&{\bar \psi}(\overset{\leftarrow}{D\ds}+m)={\bar \psi}\left(\overset{\leftarrow}{\partial}_\mu \gamma_\mu z-\frac{d}{2}\gamma_z+m\right)=0,
\eea
where $\gamma_\mu$ are the Dirac matrices of $d+1$ dimensional Euclidean space.

To solve the equations of motion, it is useful to decompose the
spinor as $\psi=\psi^++\psi^-$ where
$\psi^\pm=\frac{1}{2}(1\pm\gamma_z)\psi$. Using the equation of motion
one finds \cite{Mueck:1998iz}\footnote{ To be specific we assume
$m>0$.} \bea\label{solpsi} &&\psi^-=\int \frac{d^dk}{(2\pi)^d}\;
e^{-ik\cdot \vec{X}} z^{(d+1)/2} K_{m+\frac{1}{2}}(kz)a^-(k),\cr
&&\psi^+=\int \frac{d^dk}{(2\pi)^d}\; e^{-ik\cdot \vec{X}}
z^{(d+1)/2} K_{m-\frac{1}{2}}(kz)a^+(k), \eea where $a^{\pm}(k)$ are
arbitrary spinors satisfying $\gamma_za^\pm(k)=\pm a^\pm(k)$.
Moreover, \be a^-(k)=\frac{i \vec{k}\cdot \gamma}{k}a^+(k). \ee

To proceed, we will have to impose proper boundary conditions on the
spinors at the boundaries $Q$ and $N$. For boundary $N$, we will
follow the standard dictionary of AdS/CFT correspondence. More
precisely, for $m>0$,  using the asymptotic behavior of the modified
Bessel function near $z=0$, one finds that $\psi^-$ diverges leading
to the source term for the dual operator \be\begin{split}
\psi^-(z,\vec{X})\sim &
z^{\frac{d}{2}-m}2^{m-{1\over2}}\Gamma(m+{1\over2})\int\frac{d^dk}{(2\pi)^d}
e^{-i\vec{k}\cdot \vec{X}}\; k^{-m-\frac{1}{2}}a^-(k)+\cdots \cr
=&z^{\frac{d}{2}-m}\int\frac{d^dk}{(2\pi)^d} e^{-i\vec{k}\cdot
\vec{X}}\;\psi_0^-(k)+\cdots, \end{split}\ee where $\psi^-_0$ is the
source of the dual fermionic operator. By making use of this
notation the solutions \eqref{solpsi} read \bea
&&\psi^-={1\over2^{m-{1\over2}}\Gamma(m+{1\over2})}\int
\frac{d^dk}{(2\pi)^d}\; e^{-ik\cdot \vec{X}}
z^{(d+1)/2}\;k^{m+\frac{1}{2}}\;
K_{m+\frac{1}{2}}(kz)\psi_0^-(k),\cr
&&\psi^+={1\over2^{m-{1\over2}}\Gamma(m+{1\over2})}\int
\frac{d^dk}{(2\pi)^d}\; e^{-ik\cdot \vec{X}}
z^{(d+1)/2}\;k^{m-\frac{1}{2}}\;
K_{m-\frac{1}{2}}(kz)(-i\vec{k}\cdot \gamma)\;\psi_0^-(k).\cr &&
\eea So far we have considered the boundary condition at $z=0$. Now
we should also impose a proper boundary condition on the boundary
$Q$ at $y=0$. Following the boundary condition on boundary $N$, it
is natural to impose the Dirichlet boundary condition on $\psi^-$ at
$y=0$.  Doing so one finds \bea &&\psi^-(z,y,x_i)=\int
\frac{d^dk}{2(2\pi)^d}\; e^{-i\omega_ix_i}
{(e^{-iqy}-e^{iqy})\over2^{m-{1\over2}}\Gamma(m+{1\over2})}
z^{(d+1)/2}\;k^{m+\frac{1}{2}}\;
K_{m+\frac{1}{2}}(kz)\psi_0^-(\omega_i,q),\cr
&&\psi^+(z,y,x_i)={1\over2^{m-{1\over2}}\Gamma(m+{1\over2})}\int
\frac{d^dk}{2(2\pi)^d}\; e^{-i\omega_ix_i}
z^{(d+1)/2}\;k^{m-\frac{1}{2}}\; K_{m-\frac{1}{2}}(kz)\;\\
&&\;\;\;\;\;\;\;\;\;\;\;\;\;\;\;\;\;\;\;\;\;\;\;\;\;\;\;\;\;\;\;\;\;
\;\;\;\;\;\times
\left[-i\omega_i\gamma_i(e^{-iqy}-e^{iqy})-iq\gamma_y(e^{-iqy}+e^{iqy})\right]
\psi_0^-(\omega_i,q).\nonumber
\eea
Moreover, it turns out that the source would be an odd function with respect to $q$, {\it i.e.}
$\psi_0^-(\omega_i,-q)=-\psi_0^-(\omega_i,q)$. Here, $d^dk$ stands for $d^{d-1}\omega
dq$.

Similarly, one can solve the equation of motion for the conjugate
spinor $\bar{\psi}$. Again, it is useful to decompose the conjugate
spinor as $\bar{\psi}=\bar{\psi}^++\bar{\psi}^-$, where
$\psi^\pm=\frac{1}{2}\bar{\psi}(1\pm\gamma_z)$. Imposing the proper
boundary condition the equation of motion can be solved leading to
the following solutions:
 \bea &&\bar{\psi}^+(z,y,x_i)=\int
\frac{d^dk}{2(2\pi)^d}\; e^{-i\omega_ix_i}
{(e^{-iqy}-e^{iqy})\over2^{m-{1\over2}}\Gamma(m+{1\over2})}
z^{(d+1)/2}\;k^{m+\frac{1}{2}}\;
K_{m+\frac{1}{2}}(kz)\bar{\psi}_0^+(\omega_i,q),\cr &&{\bar
\psi}^-(z,y,x_i)={1\over2^{m-{1\over2}}\Gamma(m+{1\over2})}\int
\frac{d^dk}{2(2\pi)^d}\; e^{-i\omega_ix_i}
z^{(d+1)/2}\;k^{m-\frac{1}{2}}\; K_{m-\frac{1}{2}}(kz)\;\\
&&\;\;\;\;\;\;\;\;\;\;\;\;\;\;\;\;\;\;\;\;\;\;\;\;\;\;\;\;\;\;\;\;\;
\;\;\;\;\;\times {\bar\psi}_0^+(\omega_i,q)
\left[i\omega_i\gamma_i(e^{-iqy}-e^{iqy})+iq\gamma_y(e^{-iqy}+e^{iqy})\right],
\nonumber \eea
with the condition
$\bar{\psi}_0^+(\omega_i,-q)=-\bar{\psi}_0^+(\omega_i,q)$.

Now we have all the ingredients to compute the on-shell action. Indeed, plugging
the solution into the action, the bulk term vanishes while from the
boundary term, after adding  a proper counterterm to subtract the divergent term, one finds
\be S_{\rm on\; shell}=2A_0\int
\frac{d^dk}{(2\pi)^d} k^{2m}\;\bar{\psi}^+_0(\omega_i,q)
\;\frac{i\omega_i\gamma_i}{k}\;\psi_0^-(-\omega_i,-q), \ee
where $A_0$ is a numerical constant. Note that
in comparison with the case where there is no  boundary, in the above expression, the term
 $q\gamma_y$  is absent. We note, however, that it is
possible to add this term to the equation, though the contribution of
this term vanishes due to the fact that
 the source is an odd function with respect to $q$.  Taking this comment  into account one can
Fourier transform the source to find \bea &&\frac{2S_{\rm
on\;shell}}{A_0}=\int d^{d-1}x_1d^{d-1}x_2dy_1dy_2\int
\frac{d^dk}{(2\pi)^d}\;
e^{i\omega_i(x_1-x_2)_i}(e^{iqy_1}-e^{-iqy_1})(e^{-iqy_2}-e^{iqy_2})\cr
&&
\;\;\;\;\;\;\;\;\;\;\;\;\;\;\;\;\;\;\;\;\;\;\;\;\;\;\;\;\;\;\;\;\;\;\;\;\;\;\;\;\;\;\;\;\;\;\;\;\;\;\;
\;\;\;\times k^{2m} \bar{\psi}^+_0(\vec
x_1,y_1)\frac{i(\omega_i\gamma_i+q\gamma_y)}{k}\;\psi_0^-(\vec
x_2,y_2). \eea
which may be written as
\be S_{\rm
on\;shell}=\frac{1}{4}\bigg[I^{(0)}(y_1,y_2)-I^{(0)}(y_1,-y_2)-I^{(0)}(-y_1,y_2)+I^{(0)}(-y_1,-y_2)
\bigg].
\ee
Here $I^{(0)}(y_1,y_2)$ is the on shell action of the fermion for
the case where the space  has no boundary and is given by
\cite{Mueck:1998iz}:
\bea &&I^{(0)}(y_1,y_2)=2A_0\int
d^{d-1}x_1d^{d-1}x_2dy_1dy_2\;
e^{i\omega_i(x_1-x_2)_i+iq(y_1-y_2)}\\
&&\;\;\;\;\;\;\;\;\;\;\;\;\;\;\;\;\;\;\;\;\;\;\;\;\;\;\;\; \times
\bar{\psi}^+_0(\vec
x_1,y_1)\frac{i\gamma_i(x_1-x_2)_i+i\gamma_y(y_1-y_2)}{ |(\vec
x_1-\vec x_2)^2+(y_1-y_2)^2|^{d+2m+1}}\;\psi_0^-(\vec
x_2,y_2).\nonumber \eea
From this result, we conclude that if we
assume a coupling between the source and the dual operator as
 \be \int  d^dx\; (
\bar{\psi}^+_0\chi^++\bar{\chi}^-\psi_0^-), \ee then the two-point
function of the dual operator in the boundary CFT can be written as
a summation of four two-point functions of a CFT defined in the whole space without the
boundary as follows: \bea \langle
\chi^+(\vec{X}_1)\bar{\chi}^-(\vec{X}_2)\rangle_{BCFT}&=&\frac{1}{4}\bigg[
\langle \chi^+(\vec{X}_1)\bar{\chi}^-(\vec{X}_2)\rangle_{CFT}-
\langle \chi^+(\vec{X}_1)\bar{\chi}^-(\vec{X}^*_2)\rangle_{CFT}\cr
&&\cr &-& \langle
\chi^+(\vec{X}^*_1)\bar{\chi}^-(\vec{X}_2)\rangle_{CFT}+ \langle
\chi^+(\vec{X}^*_1)\bar{\chi}^-(\vec{X}^*_2)\rangle_{CFT}\bigg].
\eea
Using the explicit expression for the two-point function of the CFT
theory \be \langle
\chi^+(\vec{X}_1)\bar{\chi}^-(\vec{X}_2)\rangle_{CFT}\sim
\frac{\gamma\cdot
(\vec{X}_1-\vec{X}_2)}{|\vec{X}_1-\vec{X}_2|^{d+2m+1}} \ee and
taking into account that $\gamma_yq$ factor drops out of the
expression of two-point function, one arrives at
\bea
\langle
\chi^+(\vec{X}_1)\bar{\chi}^-(\vec{X}_2)\rangle_{BCFT}&=&2A_0\gamma_i(x_1-x_2)_i
\bigg(
\frac{1}{|\vec{X}_1-\vec{X}_2|^{d+2m+1}}-\frac{1}{|\vec{X}_1-\vec{X}^*_2|^{d+2m+1}}
\cr &&\cr&&-
\frac{1}{|\vec{X}^*_1-\vec{X}_2|^{d+2m+1}}+\frac{1}{|\vec{X}^*_1-\vec{X}^*_2|^{d+2m+1}}
\bigg), \eea
Utilizing the notation we have used in the previous section the
above expression can be recast into the following form \bea \langle
\chi^+(\vec{X}_1)\bar{\chi}^-(\vec{X}_2)\rangle_{BCFT}&=&\gamma_i(x_1-x_2)_i
 F(\zeta)\cr &&\cr
&\times&
\left[{|\vec X_1-\vec X\,^*_1||\vec X_2-\vec X\,^*_2|\over |\vec X_1-\vec X_2|
|\vec X\,^*_1-\vec X\,^*_2||\vec X_1-\vec X\,^*_2||\vec X\,^*_1-\vec X_2|}\right]^{\frac{d+1}{2}+m}
 \eea
where
\be
F(\zeta)=({\rm constant})\left[(\zeta+1)^{\frac{d+1}{2}+m}-\zeta^{\frac{d+1}{2}+m}\right].
\ee

\section{Conclusions}
In this letter we have explored some aspects of AdS/BCFT correspondence, including
 the asymptotic symmetry of geometries which are conjectured to provide the
holographic dual of BCFTs.  In particular, we have demonstrated that
in the two-dimensional BCFT, the corresponding asymptotic symmetry of the
dual geometry is indeed two copies of the Virasoro algebra subject
to a constraint relating the left and right moving generators.

Using the general dictionary of AdS/CFT correspondence, we have also
computed two-point functions of certain bosonic and fermionic
operators in a BCFT, by making use of its holographic dual. The
resultant correlation functions are in agreement with those in the
literature of BCFT. Therefore, our results may be considered as a
check for the newly proposed AdS/BCFT correspondence.

It should be mentioned that in our study we have considered the
simplest examples in which the BCFT lives in half space (upper half
plane in two dimensions).  From the bulk theory point of view, the
corresponding  gravitational theory lives in a portion of AdS
geometry space separated from other parts by a simple hypersurface
given by  $y=0$. This is the hypersurface where  the metric
satisfies the Neumann boundary condition.

We have observed that in this simple holographic model, the two-point
functions of the operators have a symmetric structure reminiscent of
method of image in electrostatic. More precisely, the two-point function
of the BCFT in half space can be written in terms of four two-point
functions of the operators and their images in a CFT which is defined
in whole space without any boundary.

Generalization to more complicated boundaries is straightforward,
though a little bit tedious. In particular, one may consider a BCFT
on a disc or strip.  One would expect that in these cases, the method
of image can be also used to write the corresponding two-point
function though, the procedure is more involved.


\section*{ Acknowledgments}

We would like to thank Ali Naseh for collaboration in the early
stage of this project, as well as useful discussions. We would also
like to thank  Davod Allahbakhshi, Mohammad R. Mohammadi,  Ali Mollabashi and
M. M Sheikh Jabbari for useful
discussions.


\begin{thebibliography}{99}

\bibitem{Maldacena:1997re}
  J.~M.~Maldacena,
  ``The Large N limit of superconformal field theories and supergravity,''
  Adv.\ Theor.\ Math.\ Phys.\  {\bf 2}, 231 (1998)
  [Int.\ J.\ Theor.\ Phys.\  {\bf 38}, 1113 (1999)]
  [arXiv:hep-th/9711200].

\bibitem{Gubser:1998bc}
  S.~S.~Gubser, I.~R.~Klebanov and A.~M.~Polyakov,
  ``Gauge theory correlators from noncritical string theory,''
  Phys.\ Lett.\  B {\bf 428}, 105 (1998)
  [arXiv:hep-th/9802109].

\bibitem{Witten:1998qj}
  E.~Witten,
  ``Anti-de Sitter space and holography,''
  Adv.\ Theor.\ Math.\ Phys.\  {\bf 2}, 253 (1998)
  [arXiv:hep-th/9802150].


\bibitem{Takayanagi:2011zk}
  T.~Takayanagi,
  ``Holographic Dual of BCFT,''
  arXiv:1105.5165 [hep-th].

\bibitem{Fujita:2011fp}
  M.~Fujita, T.~Takayanagi and E.~Tonni,
  ``Aspects of AdS/BCFT,''
  arXiv:1108.5152 [hep-th].



\bibitem{Gibbons:1976ue}
  G.~W.~Gibbons and S.~W.~Hawking,
  ``Action Integrals and Partition Functions in Quantum Gravity,''
  Phys.\ Rev.\  D {\bf 15}, 2752 (1977).






\bibitem{Brown:1986nw}
  J.~D.~Brown, M.~Henneaux,
  ``Central Charges in the Canonical Realization of Asymptotic Symmetries: An Example from Three-Dimensional Gravity,''
  Commun.\ Math.\ Phys.\  {\bf 104}, 207-226 (1986).

\bibitem{Bachas:2002nz}
  C.~Bachas,
  ``Asymptotic symmetries of AdS$_2$-branes,''
  arXiv:hep-th/0205115.



\bibitem{Skenderis:2002wp}
  K.~Skenderis,
  ``Lecture notes on holographic renormalization,''
  Class.\ Quant.\ Grav.\  {\bf 19}, 5849 (2002)
  [arXiv:hep-th/0209067].

\bibitem{Cardy:1984bb}
  J.~L.~Cardy,
  ``Conformal Invariance and Surface Critical Behavior,''
  Nucl.\ Phys.\  B {\bf 240}, 514 (1984).

\bibitem{Henneaux:1998ch}
  M.~Henneaux,
  ``Boundary terms in the AdS / CFT correspondence for spinor fields,''
  arXiv:hep-th/9902137.

\bibitem{Henningson:1998cd}
  M.~Henningson and K.~Sfetsos,
  ``Spinors and the AdS / CFT correspondence,''
  Phys.\ Lett.\  B {\bf 431}, 63 (1998)
  [arXiv:hep-th/9803251].

\bibitem{Mueck:1998iz}
  W.~Mueck and K.~S.~Viswanathan,
  ``Conformal field theory correlators from classical field theory on anti-de
  Sitter space. 2. Vector and spinor fields,''
  Phys.\ Rev.\  D {\bf 58}, 106006 (1998)
  [arXiv:hep-th/9805145].



\bibitem{Laia:2011zn}
  J.~N.~Laia and D.~Tong,
  ``A Holographic Flat Band,''
  arXiv:1108.1381 [hep-th].






\end{thebibliography}
\end{document}